\documentclass[twocolumn]{article} 
\usepackage{graphicx} 
\usepackage{mathtools}
\usepackage[numbers,square]{natbib} 
\usepackage[english]{babel}
\usepackage{url}

\long\def\comment#1{}

\makeatletter
\def\@normalsize{\@setsize\normalsize{10pt}\xpt\@xpt
\abovedisplayskip 10pt plus2pt minus5pt\belowdisplayskip 
\abovedisplayskip \abovedisplayshortskip \z@ 
plus3pt\belowdisplayshortskip 6pt plus3pt 
minus3pt\let\@listi\@listI}

\def\subsize{\@setsize\subsize{12pt}\xipt\@xipt}
\def\section{\@startsection {section}{1}{\z@}{1.0ex plus
1ex minus .2ex}{.2ex plus .2ex}{\large\bf}}
\def\subsection{\@startsection 
   {subsection}{2}{\z@}{.2ex plus 1ex} {.2ex plus .2ex}{\subsize\bf}}
\makeatother

\begin{document}

\date{}

\title{\Large {\bf Event based classification of Web 2.0 text streams}}

\author{Andreas Bauer\\
Media Computer Science\\
University of Regensburg\\
  \texttt{andreas.bauer@extern.uni-regensburg.de}
\and
Christian Wolff\\
Media Computer Science\\
University of Regensburg\\
 \texttt{christian.wolff@sprachlit.uni-regensburg.de}
}

\maketitle
\thispagestyle{empty}

\subsection*{\centering Abstract}
{\em 
Web 2.0 applications like Twitter or Facebook create a continuous stream of information. This demands new ways of analysis in order to offer insight into this stream right at the moment of the creation of the information, because lots of this data is only relevant within a short period of time. To address this problem real time search engines have recently received increased attention. They take into account the continuous flow of information differently than traditional web search by incorporating temporal and social features, that describe the context of the information during its creation. Standard approaches where data first get stored and then is processed from a peristent storage suffer from latency. We want to address the fluent and rapid nature of text stream by providing an event based approach that analyses directly the stream of information. In a first step we want to define the difference between real time search and traditional search to clarify the demands in modern text filtering. In a second step we want to show how event based features can be used to support the tasks of real time search engines. Using the example of Twitter we present in this paper a way how to combine an event based approach with text mining and information filtering concepts in order to classify incoming information based on stream features. We calculate stream dependant features and feed them into a neural network in order to classify the text streams. We show the separative capabilities of event based features a base for a real time search engine. 

Keywords: 
information retrieval, text mining, event processing, web2.0, text streams, real time search, neural network, stream features
}


\section{Introduction}
\label{Introduction}

Instantaneous information sharing offered by services like Twitter, Facebook or Tumblr led to a major increase of user generated content. This creates a continuous stream of information which poses new challenges regarding processing, analysis, information retrieval and filtering. While traditional search engines like Google, Bing or Yahoo are focused on delivering information which covers the whole range of relevants facts regarding a given search query, real time search engines like new Google+, Twitter Search or topsy.com intend to deliver insight into the continuous information stream right at the moment when the information is created, i.e. the tolerance for latency compared to traditional search engines is very low\cite{Efron:wh}.

In this paper real time search engines are considered as being a system that removes irrelevant items from a continuous stream of data. Hence real-time search engines are very similar to information filtering systems as described in \cite{Belkin:1992tx}. This type of real-time search engine should not be confused with such search engines that deliver results instantaneously from a given after the query has been submitted to the system, e. g. Solr. 

In this paper we want to describe the basis for an event based information filtering system that analyses real time text streams like the ones produces by Twitter or Facebook. Hence our approach can be considered as the basis for a real time search engine using event processing methods.  The contributions are made with this paper

\begin{itemize}
  \setlength{\itemsep}{1pt}
  \setlength{\parskip}{0pt}
  \setlength{\parsep}{0pt}
\item Introduction of atomic information events
\item Mapping of text streams to different information event types
\item Calculation of stream based event features 
\item Training and evaluation of neural networks with stream based features
\item Evaluation of the performance of the trained networks
\label{contributions}
\end{itemize}

The paper is structured as follows. In chapter 2 we give an overview of the already existing literature and research. Chapter 3 defines the problem definition, while chapter 4 provides the theoretical basis. In chapter 5 we show the application and evaluate our approach.

\section{Related work}
\label{Related work}

This paper combines several research areas. \cite{Belkin:1992tx} provides the basic framework for our research. The paper defines the basics of information filtering and introduces basic concepts we also use in this paper, e.g. analysis of text streams, standing queries, removal of information from the stream, user profiles, etc. \cite{Haghani:2010p1996} introduce a way of real-time, top-k and profile based information filtering with sliding time windows based on the traditional $tf.idf$ weighting method. The goal and approach of their paper is similar to this paper's goal. Our approach differs as we use neural networks to learn measures of interest and use stream based features for calculating similarity measures. ``Topic Detection and Tracking'' addresses similar questions like information filtering. Noteworthy in the context of Twitter topic detection is \cite{Cataldi:2010p1039}. They use an ageing theory based approach for judging new Tweets. \cite{CheAlhadi2012LMA} also use a similar set-up for detecting interesting Tweets. But first they use incremental Naive-Bayes-Classifier and second they do not focus on the usage of events and stream features.

Stream processing and data stream mining is also closely related to the topics of this paper. \cite{Bifet:2010p985} describe a data streaming approach for sentiment analysis in Twitter data. They apply multinomial na\"ive Bayes, stochastic gradient descent and Hoeffding trees in order to identify sentiments from Tweets. In contrast to this paper we focus on filtering Tweets, we use neural networks and in our set-up the trained networks are only applied to a short sliding time window and then are rebuilt from scratch. \cite{Guc:2010p1490} combines incremental decision trees and streamed text features in order to filter Tweets from Twitter lists. It also includes real time aspects as it uses the Twitter Streaming API, but its goal mainly to filter Tweets from group lists with less frequent changes.

Real time search and real time information filtering is addressed in \cite{Phelan:2009p443}, who describe a system where they use current information available about the Twitter stream and the user, in order to build an information retrieval system based on $tf.idf$, which filters tweets from an incrementally growing tweet corpus. Also \cite{Mathioudakis:2010fc} introduces a way using trending detection within the Twitter stream in order to filter irrelevant information. The microblog track from the TREC 2011 conference is also related, but this track focused on the retrieval of Tweets from a non-dynamic corpus. The task was also called real time retrieval, but more in the sense of finding all relevant Tweets up to a given timestamp. This contrast from our approach as we focus on real time filtering. Also related to this paper is \cite{Duan:2010wi} who propose a learning to rank approach for Tweets. This is similar in terms of feature engineering and machine learning. But we use a rather dynamic, continuous way of learning and we use an artificial target function, while they propose the analysis of a static Twitter corpus with manual annotated gold standard. 

\section{Problem definition}
\label{Problem definition}

\begin{figure}[ht]
  \centering
  \includegraphics[scale=0.30]{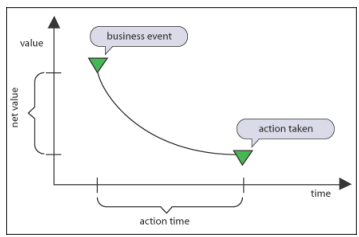}
  \caption{value time curve \cite{hackathorn2004}}
  \label{images/value_time_curve_event}
\end{figure}

The continuous flow of information from Web 2.0 sources opens the application for a new type of search engines that deliver results based on content that gets generated during the information seeking episode and are not based only on a periodically updated document corpus like in classic search engines. The so called real time search engines deal with a continuously changing document corpus and address the problem that they permanently have to rank and classify incoming new items and present the updated result immediately to the user. This is in contrast to classic web search where the ranking of a document is precomputed in some way based on a given relevance schema and a static corpus size at the time of computation. Of course current web search engines update their corpus many times each second and constantly recalculate the metrics for their search engine, but this applies to their overall corpus, i.e. the overall corpus is growing each second but it's not assured that the corpus regarding a real time search need is also growing every second. Regarding a real time search episode the content crawled by classic search engines does not update that frequently thus only the aforementioned Web 2.0 stream can provide new documents for the corpus of the real time search. 
Real time search engines used a \textit{micro corpus} which is being updated continuously and is only valid a certain amount of time into the past. The fact that classic web search engines blend real time search results into their classic web search results shows the importance of real time content. \ref{images/value_time_curve_event} shows the decay in relevance of information events over time. Hence a processing close to the creation of an information event is sensible.

Our research focus in on the continuous removal of irrelevant data from a text stream. The information filtering episode we consider is that a user is interested in a certain topic. It is not given that the user has an elaborated personal web profile or if she has one then it might not be related to the information filtering episode. E.g. the user mainly tweets about machine learning and data mining, but for her information filtering episode she is interested in the US pre-elections. Thus her personal Twitter profile won't necessarily match.

While the interaction in the Web 2.0 entails not only text, but also multimedia, we focus only on textual information streams in this paper. The properties of a continuous text stream is the main challenge for real-time search engines and demands special attention. We focus on the Twitter stream due to its availability, but in later research we want to apply our approach also to other text streams. 

Characteristics of Web 2.0 text streams based on results from \cite{Dong:2010p1686} and \cite{teevan:twittersearch} show the peculiarities, which have to be considered.

\begin{itemize}
  \setlength{\itemsep}{1pt}
  \setlength{\parskip}{0pt}
  \setlength{\parsep}{0pt}
\item Recency
\item Timeliness
\item Data Volume
\item Temporal Validity
\item Temporal relevance
\item Social interaction
\item Shortness
\item Dynamic corpora (terms, documents)
\label{textstreamprops}
\end{itemize}

To our knowledge only few information is publicly available on how real time search engines assess incoming information. Google's real-time search - before it was closed down in July 2011  and later incorporated into Google+ - was supposed to use an algorithm similar to PageRank, where not the link structure was taken into account, but the reputation of the user \cite{talbot:googletweet}. http://www.topsy.com also relies on a influence factor of the content creating users. Other search engines like SocialMention.com, kurrently.com or 48ers seem to take into account key word filtering as well as social features, but nothing is known about their approaches. The measure of all real-time search is Twitter itself. Their corpus is of course sublime, as they have all the Twitter data, but we think our approach adds an interesting aspect to this topic.

As a summary table \ref{tab:rt_classic} summarizes the main differences between real time and classic web search.

\begin{table}
\begin{tabular}[ht]{{|p{2,2cm}|p{2,5cm}|p{2,2cm}|}}
\hline
Aspect & Real time search & classic search\\
\hline
Corpus & dynamic & static\\
\hline
Corpus update & continuous & periodic\\ 
\hline
temporal relevance & short & arbitrary\\
\hline
Query modification & rare & frequent\\
\hline
Content generation & spontaneous, ad-hoc & elaborated\\
\hline
Document length & short  & arbitrary \\
\hline
\end{tabular}
\caption{\label{tab:rt_classic}Comparison features real time vs. classic web search \cite{teevan:twittersearch}}
\end{table}

In order to add some new ideas to this interesting and rising topic we like to introduce an intuitive approach for ranking real-time content, which combines document features as well as term weighting with text mining and event processing. While the first three aspects are well established in the areas of information retrieval and ranking, the latter allows to react on real time content and to determine features for incoming items as they occur. The combination of these four research areas allow us to address the problems that are posed by the characteristics of Web 2.0 text streams \ref{textstreamprops}. 

We want to introduce an event based approach to tackle these problems. This approach uses methodologies from event processing which apply naturally to the demands of streamed text data. The advantage of event processing is that you do not have to build temporal data structures and window based calculations yourself, but you can use sophisticated stream engines which offer advanced functionalities like sliding windows or pattern matching. We combine this with the foundations of information retrieval and information filtering to answer questions 1-4.

There are many research papers on analysing (\cite{Golder:2010p434}, \cite{Java:2007p360}), classifying (\cite{Guc:2010p1490},\cite{Bifet:2010p985}) or ranking (TunkRank, etc.) with continuous text streams - foremost Twitter - , but all have in common that a several human estimators had to provide classes or rankings in order to train the models. We try to overcome this task by using stream properties for training the ranking function and for ranking the text stream snippets. This is foremost important as our research focus lies on ad-hoc, short-term filtering and monitoring tasks for Web 2.0 text streams where we assume that during the time of the information episode no elaborated user preference model is available. 
This is in contrast to user profile based search or collaborative search, where you can use historical data of the user itself or a group of users in order to build a user model that can be used in the search. We exclude this fact in this paper and save the analysis for later research. In this paper we assume that it is not always possible to include a user profile because either the profile is simply not available, not yet elaborated enough to be taken into account or is not related to the filtering task. Hence the proposed approach shows one component for a real time information filtering systems. The query matching part also will be presented in another research paper.

\section{Event based information filtering for continuous web 2.0 text streams}
\label{Event based information filtering and retrieval for continuous web 2.0 text streams}

In the next section we want to introduce the basics for an event based information filtering system for real-time text streams from web 2.0 sources, show why an event driven approach is well suited for this. 

In the first part of this chapter we show how to map a raw tweet onto several independent event types, which in turn are fed into individual streams for the analysis. In the second step we show how you can apply quantitative and qualitative stream filtering and pattern matching methods in order to extract relevant events from the several streams. Then in the third part we show how we use these filtered events for ranking single events within the stream and how we use this approach for classifying tweets into several categories. The evaluation and examination follows in section 6.

\paragraph{Event based information filtering}
\label{event definition}

Before we start to explain our event based system, we want to clarify the notion of the term \textit{event} we use through out this paper. 
While \textit{event} is a very generic term and is overloaded with different meanings in different research areas, we use the notion coming from the research area of \textit{event processing systems}. This foremost based on the definition of \cite{OpherEtzion:2010tv} and \cite{Luckham:2002tu} and is already widely used in areas of logistics or financial services. We want to introduce the concept of atomic information events for information retrieval and filtering.

In our approach we consider the text stream information as consisting of smaller, simpler events, i.e. each tweet, Blog post or Facebook update is made up of several tinier events that belong to different event classes like token events, location events or link events. This allows us to analyse not only the stream of incoming ''raw events'' like Tweets, but also we can analyse the underlying ''simple'' events that are the building blocks of the raw event and combine them to more complex events. 

One might argue that using the term \textit{event} instead of the term \textit{token} does not have any relevance for text mining and information filtering/retrieval. But we emphasize that the use of the event metaphor offers several advantages. 
First the term event is a semantic construct, which underpins the temporal and dynamic character of data stream. It suitable to describe the state of information from their time being created to the time it makes its way to \textit{permanent} part of the web. During this time the information event's signal is the strongest and offers the most potential for analysis. 
Second for a real-time filtering system it is sensible to only consider a limited time horizon as relevant. Hence events that leave this horizon have lost the majority of their importance. This is supported by the temporal semantics implied with events. 
Third events are only evaluated once within an event processing network, i.e. after they have entered the network they run  through all standing queries and patterns that are in place and then leave the processing work flow. This supports instantaneous processing of the data stream. 
Fourth an event is not bound to a distinct processing step. The system can be flexibly extended as events are not bound to a specific event processing agent. Last the event based approach offers the ability of processing information right in its natural and actual context. The context for information, that has already been persisted, has to be artificially recreated, i.e. there is a notable amount of overhead for putting the information in its original state. Thus the usage of information events is highly efficient compared to persisted information items and allows the creation of executable knowledge right at the creation time of information.

\subsection{Event mapping and stream creation}
\label{Event mapping and stream creation}

First we build several  independent streams from the raw torrent of tweets. For this we process the structured meta data part of the tweet and the un- and semi-structured part of the tweet separately. The metadata information such as location, followers or friend count, number of tweets or local time can be mapped directly to distinct event types. These events are all composed of the following parts \cite{OpherEtzion:2010tv}:
\begin{itemize}
  \setlength{\itemsep}{1pt}
  \setlength{\parskip}{0pt}
  \setlength{\parsep}{0pt}
\item header / description attributes: 
\subitem unique event identifier
\subitem timestamp and time granularity
\subitem event source
\item payload: actual event information
\end{itemize}

For our event based information filtering/text analysis system we map a tweet onto the following event types

\begin{itemize}
  \setlength{\itemsep}{1pt}
  \setlength{\parskip}{0pt}
  \setlength{\parsep}{0pt}
\item Token score event
\item Link events
\item Retweets events 
\item Hashtags events
\item Cooccurrence events
\item Metadata events: status, follower count, followee count
\end{itemize}

The mapping process of a tweet is split into two parts and is performed by Event Processing Agents (EPA) \cite{OpherEtzion:2010tv}, which are placed in the text stream. The one for the structured, directly accessible meta data part, i.e. information that can be directly extracted from the tweet without further processing. In theory you could spare this step and operate directly on the properties of the raw event, i.e. the tweet. But in order to have clear semantics which support and ease the definition of the processing logic for the Twitter stream it is recommended to map the information on separate event types. Additionally this offers the advantage that you can directly operate on the event information without additional filtering for the desired attribute and second you avoid unnecessary payload overhead, which can have impacts on performance when you take into account that several thousand events per second are processed.  

The second step deals with extracting and mapping of the un- and semi-structured text part of a tweet. In this case un- or semi-structured means, that the information within the text itself has to be made processable for an event processing engines. In order to access the information buried within the text, we apply on the one hand standard text mining preprocessing and on the other hand we extract additional semantics from the text based on the characteristics of tweets.
The first step is used for creating linguistic base event types like $TokenEvent$ or $CooccurrenceEvent$ and includes tokenization, stemming and stop word removal. At the end of this process we get a normalized token which you can use to populate, e.g. a $TokenEvent$, with its payload. 

The second step uses known semantics of Tweets that were introduced by Twitter or Twitter users to provide additional information. This includes mentions (@ sign), hashtags (\#), retweets and links. We use regular expressions in order to extract the information. Each encountered item is mapped onto the corresponding event type. The aforementioned semantics are typical for Twitter, but this approach can be extended and used for every event source that provides extractable semantic features.

At the end of the process we end up with several independent information streams which can be analysed separately, but can also be joined by their common event attributes, e.g. the unique of the tweet the events were derived from, or finally employed for event pattern matching.

\subsection{Using stream properties and event patterns for ranking and classification}
\label{Using stream properties and event patterns for ranking and classification}

In this section we want to describe how to use stream features for real time classification purposes.

In our analysis we employ event processing techniques to select and build features for the machine learning algorithm. The features are then fed into an artificial neural network (ANN). \cite{Govindarajan:vj} gives an overview of the advantages of using neural networks for text classification. We use a neural network because they are well suited for pattern detection, are capable to deal with dynamic and incomplete data. All three properties are typical for our set-up, as we look for patterns in a sampled and dynamic text stream.

After the network is trained the new events, that have arrived within a defined sliding time window, are assessed and the top percentile is kept for further evaluation. The results of this step can be considered again as new events that can be fed into other event processing agents. Combined with events from EPAs, that deal with query evaluation or context definition, the information events can be combined to a final result that is presented to the user. But to keep the focus clear for this paper, we concentrate on the description of the basics of the event based systems and one of its application - the usage for training machine learning algorithms.

\subsubsection{Defining quantitative measures for text stream analysis}

By mapping the features of a tweet onto separate event types and feeding the events into separate streams, we now can directly operate on the events with established stream processing methods. These entail the calculation of stream statistics on sliding windows, detecting patterns like drops and burst, recognition of event sequences and correlating of different streams.  
Furthermore we can use the capabilities of stream engines to provide statistical information (standard deviation, variance) on different properties of a stream. Combining these features we are able to create a focused research corridor which allows us to judge Twitter messages as they occur. In this paper we mainly focus on volume based features.

\paragraph{Stream characteristics} An important factor for any filtering system is its real world applicability. In our experimental set-up we used an event frequency of 100 Tweets per seconds. The system is implemented in Java and uses a maximum heap size of 4GB. The experiments were conducted on a Mac with a Dual 3.2 GHz processor. Throughout the experiments the event speed could be kept constant. Besides the training time of the neural network, the delay for processing the incoming events, i.e. the time the event processing agent needed to map the stream features onto training features, was about 2 seconds.

Twitter in reality processes approx. 4000 Events per seconds. At peek times it is about twice as much. In general approx. 40\% of all Tweets are English. Hence a system has to process 1600 to 3200 Tweets per seconds. As we only had access to the Twitter garden hose with 10\% of all Tweets and as only 40\% of those Tweets are English, we chose to use an event speed of 100 Tweets per second. For testing purposes the system was also able to go up to 500 Tweets per second in its basic set-up, i.e. the multi-threading capabilities of the stream engine were not fully exploited yet. For a real-world application the performance has to be improved to process all incoming events in real-time as well as adapt to the changing amount of data. But to show the principles and the general approach this performance is sufficient.

\paragraph{Features}

In order to learn a model which can be used for classification, we have to select and calculate appropriate features. The possible features that can be used are already limited by the characteristics of a single tweet. Thus our features do not fundamentally differ from those, e.g. used by \cite{Naveed:2011vn}. 

The features are not calculated on a static document corpus but are based on sliding time windows, i.e. all features reflect their state within the defined sliding time window, e.g. if we define a sliding time window length of 30 seconds the features represent their state within those 30 seconds. This additionally covers problems with concept and topic drift within a stream as we steadily adapt to the new stream characteristics. 

We have \textit{event specific features}, like hashtag or link presence and bursty features count, which are directly dependent on the text of the event, as well as \textit{stream specific features} like token score. The latter are not directly calculated based on the event itself, but depend on the stream state for the entirety of a given event type, i.e. the token score of a single term within a tweet is derived from the stream state of that token within the sliding time window. In a textual analysis like this the score of a token depends on the occurrences of this token within the considered sliding time window. 

\begin{itemize}
  \setlength{\itemsep}{1pt}
  \setlength{\parskip}{0pt}
  \setlength{\parsep}{0pt}
\item Score for 5 token within a tweet
\item Scale normalized score for follower, friend and status count
\item Presence of link
\item Presence of hashtag
\item Frequency variation for 5 token (compared to window two minutes ago)
\label{ref:features}
\end{itemize}

\subsection{Neural network set-up and topology}

The next step in our approach is to feed the features into an artificial neural network (ANN) in order to learn the target function for the current inspected time slot. 

For this we first have to define a target function which the neural network can learn. As it is usually impossible to provide an adequate amount of labelled data for classification in the desired time frame of only a few seconds, we chose an artificial target value, which can be defined for every single tweet. We also prefer this approach, as in reality you cannot rely on extensive user labelled data (unless you are Google). In this set-up we use the presence of a Retweet of a tweet within a sliding time window of 120 seconds, i.e. the tweet itself and its Retweet occur within the aforementioned time window. The interpretation of a Retweet being a measure of interestingness is also supported by \cite{Naveed:2011vn}, \cite{Suh:2010uw} and \cite{Naveed:2011uy}. 

After we have trained an classifier for the analysed time window, we apply the classifier to new Tweets. In addition the new Tweets are used to train a new classifier from scratch. With this procedure we can provide a continuous classification of incoming Tweets.

The size of the input layer of the neural network is defined by the number of features described in the section above. We use a three layer neural network where the input layer consists of 15 input features, a hidden layer with 10 neurons and an output layer with a single neuron. The size of the hidden layer was chosen by taking the average of the output of an incremental pruning approach, i.e. for identifying a reasonable size of the hidden layer several several runs were conducted which included a pruning step. The learning algorithm is resilient back-propagation, the activation function is  function for the hidden layer and SoftMax for the output layer. The latter is due to the fact that we want to treat the outcome as the posterior probability of the tweet belonging to a class. The error function is linear as this fits well to a SoftMax activation function and the penalty calculation is based on cross entropy \cite{Bishop:1996wd}. We used the Encog Neural Network library.\footnote{\url{http://www.heatonresearch.com/encog}}.

Overview of the employed features.

\begin{itemize}
  \setlength{\itemsep}{1pt}
  \setlength{\parskip}{0pt}
  \setlength{\parsep}{0pt}
\item Hashtag indiciator (binary)
\item Link indicator (binary)
\item Score of top 4 tokens within a tweet (numeric)
\item Frequency variation of sliding time window for top 4 tokens
\item Scale normalized amount of Tweets by user
\item Scale normalized amount of friends by user
\item Scale normalized amount of followers by user
\item Scale normalized length of tweet
\label{ref:features}
\end{itemize}

The output layer has the size of one, as we deal with an information filtering problem where it is the main goal to divide  the value of a piece of information into categories of relevance and non-relevance for further processing. So with the target function defined above we labelled data as relevant or not relevant and use this information for training. 

The training and ranking for sliding time windows will work as follows: 

\begin{enumerate}
  \setlength{\itemsep}{1pt}
  \setlength{\parskip}{0pt}
  \setlength{\parsep}{0pt}
\item Define a sliding window size \textit{length($t_{i}$)}
\item Calculate streams features for sliding window $t_{i}$
\item Normalize each features by calculating the scale normalized value
\item Define a target value for each tweet (e.g. Retweet as as sign of significance and interestingness)
\item Setup a new ANN for sliding window$t_{i}$
\item Split the captured events of sliding window$t_{i}$ into test and training test
\item Train, cross-validate and test the ANN  
\item Calculate a score for each captured event of sliding window $t_{i}$ using the trained ANN
\item Return a sorted list of the scored events
\item Combine the scored events with query in order to filter the relevant tweets
\end{enumerate}

\begin{figure*}[ht]
  \centering
  \includegraphics[scale=0.30]{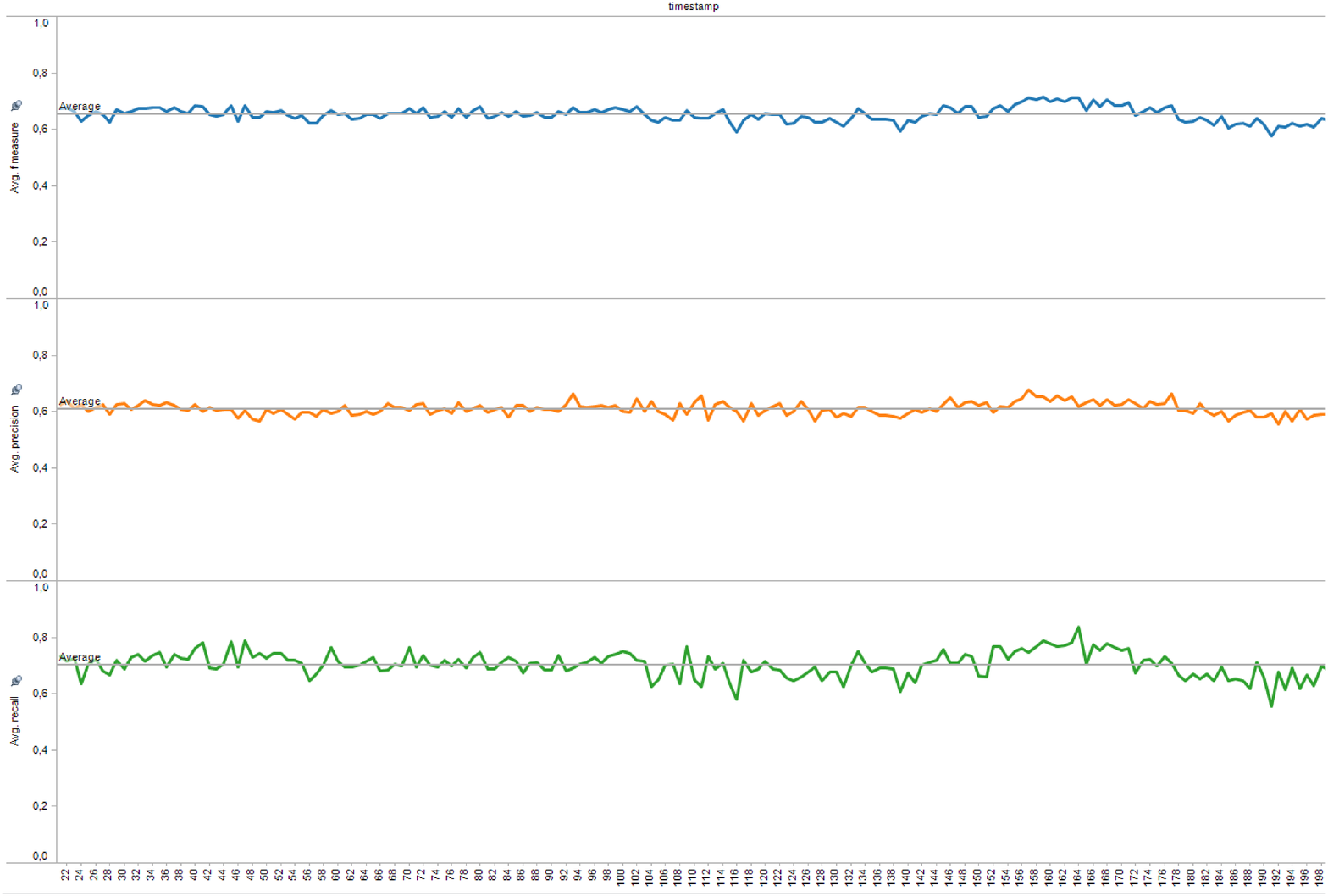}
  \caption{Measure for experiments with real data}
  \label{images/real_data}
\end{figure*}

\begin{figure*}[ht]
  \centering
  \includegraphics[scale=0.30]{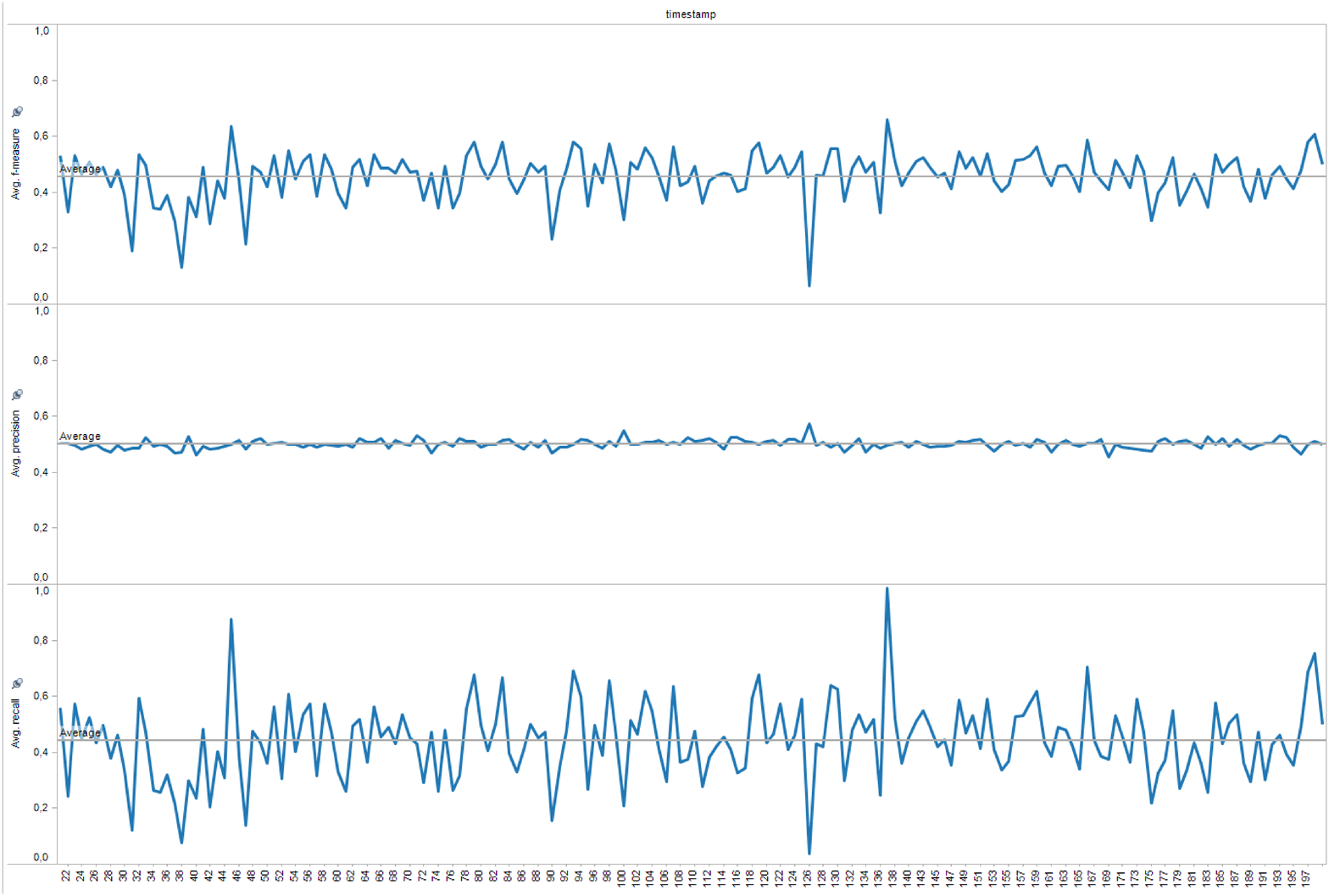}
  \caption{Measures for random experiment}
  \label{images/random_data}
\end{figure*}

\begin{table}
\begin{tabular}[ht]{{|p{2,2cm}|p{2,5cm}|p{2,2cm}|}}
\hline
Measure & Random & Real world data\\
\hline
f-measure avg. & 0.4556 & 0.6515\\
\hline
precision avg. & 0.4991 & 0.6085\\ 
\hline
recall avg. & 0.4396 & 0.7024\\
\hline
Number of observed sliding windows & 20-200 & 20-200\\
\hline
Training samples avg. & 8147 & 9917 \\
\hline
\end{tabular}
\caption{\label{tab:experiments}Results of experiments}
\end{table}

\section{Experiments and Evaluation}

In order to show the applicability of our approach we evaluate the method described in the section above on selfsampled Twitter data. 

We show the f-measures, precision and recall curves for our set-up over an information filtering episode. For this paper we chose a corpus of Tweets collected on March 2nd 2012 8.00 - 18.00 CET. The corpus contains 839.095 Tweets that we collected with keyword tracking from the gardenhose streaming API of Twitter. For repeatability purposes we saved the Tweets to a database. Furthermore we ran a language detection algorithm \cite{nakatani2010langdetect} on it, as we only want to use English Tweets.     

The whole set-up works in a streamed fashion due to its natural event based character. We defined a sliding time window of 120 seconds. To stabilize the stream the output of the sliding window will be fed into the neural network every 10 seconds. The neural network gets trained from scratch for every sliding time window again. Thus our approach uses micro-batches as neural networks that use back-propagation can not be updated continuously like e.g. Na\"ive Bayer classifier.
The ratio of positives Tweets (Tweets that were retweeted within the sliding time window) and negative Tweets is very skewed (approx. 2\%:98\%), therefore we use oversampling in order to overcome the imbalance in the dataset. \cite{Weiss:2007um} showed that this approach works despite its potential drawbacks like overfitting. To avoid overfitting we split the samples coming from the sliding time window into three independent sets, one for training, one for testing and one validation. We used a 5-fold cross validation approach and one fifth of the data was used for the testing of the trained model. Additionally we apply early stopping of the neural network after two increase in a row of the training error. And finally we set a maximum training time of 10 seconds. All measures are intended to keep a good performance for our set-up and within the real time constraints of the setting. A summary of the result can be found in table \ref{tab:experiments}

Finally we use the trained network to evaluate newly arriving Tweets. The outcome of the neural network for each new Tweets can be considered as the probability of the new tweet being similar to the pattern of the training set. We do not consider the outcome as the probability of a tweet being retweeted, but as the similarity of a tweet to another tweet that has been retweeted and thus has been considered interesting. This result can be the input for further processing with the event processing network and hence can be combined with results from other event processing agents.

To have a reasonable comparison of the performance of the system, we show the f-measures, precision and recall of our set-up compared to the measures of naive random approach, where the target values are the same like in the runs with real data, but the features were initialized randomly on scale 0 to 1. \ref{images/real_data} shows that our approach performs better than the random set-up \ref{images/random_data}. Even though we have to mention that the training and test errors between our system and the randomized approach not as distinct as the f-measure, it can be proved that the selected features are appropriate for the selected task. 
Furthermore it becomes obvious that our approach offers a stable output performance over the evaluated time frame (we only depict the results of sliding window number 20-200, in order to keep the graphs comprehensible). In \ref{images/random_data} you can see that precision values are almost exactly 0.5 what you can expect by guessing a binary classification. The recall values of the random approach differ extremely compared which underpins the assumption that there is a regular pattern within the stream that can be learned. This can be proved by the higher and more regular values produced from our approach.

The last indicator that our approach works is the Cohen kappa statistics. Over the recorded sliding time windows the kappa value for our approach was approx. 0.4, which indicates according to \cite{Landis:1977vp} a fair to moderate classification performance. To verify the set-up the kappa for the randomized data had an expected kappa statistics average of 0.

\section{Conclusions and Future Work}

In this paper we presented a event based approach for training an artificial neural network with information events. We showed how to map a tweet on different event types, analyse these information events with event processing techniques and how to combine this with machine learning approaches. We showed how to calculate the features for the neural network and evaluated the described set-up using standard performance criteria. 

It could be shown that despite being a very noisy media the information from the Twitter stream can be used to train a neural network model which gives reasonable f-measures. Even though the kappa statistics values are only fair to moderate, our approach gives good indicators for being usable in an event based filtering system. In the future we want to elaborate the distinctiveness of different feature set, apply other learning algorithms like support vector machines or online learning algorithms. Besides this \textit{computational approach} we want to investigate the heuristic based pattern approach for the analysis of information events, i.e. we want to build a net of standing queries that look for interesting events within an information stream. 

In this scenario we selected a measure of potential interestingness, but the target function is not limited to this scenario. The experiments showed the separative power of our event based features. Thus we will try different target functions, e.g. to detect spam.
Finally as this method gives only a relevance estimate for all Tweets within the stream, we want to incorporate further filtering methods in order to adapt the system to the user and deliver only information events that are  relevant for a particular user. We also intend to enlarge this approach to more elaborate web content like blogs or news feeds


\nocite{*}
\bibliographystyle{plainnat}
\bibliography{event_driven_ir}

\end{document}